\documentclass[10pt,a4paper]{article}
\usepackage[english]{babel}
\usepackage{bm,amsmath,amssymb}
\usepackage{mathpazo}
\usepackage{courier}
\usepackage{a4wide}
\usepackage{graphicx}
\usepackage{subfigure}
\usepackage{units}
\usepackage{xspace}
\usepackage{multirow}
\usepackage{appendix}
\usepackage{notations}
\usepackage{color}
\definecolor{darkblue}{rgb}{0,0,0.5}
\definecolor{firebrick}{rgb}{0.75,0.125,0.125}
\definecolor{darkgreen}{rgb}{0,0.5,0}
\definecolor{light-gray}{gray}{0.5}
\usepackage[colorlinks=true,linkcolor=firebrick,citecolor=darkgreen,urlcolor=darkblue]{hyperref}

\usepackage{natbib}
\usepackage[textwidth=2.4cm, textsize=scriptsize]{todonotes}

\def\HRule{\rule{\linewidth}{1mm}}

\def\b0{b_0}

\begin{document}

\thispagestyle{empty}
\begin{flushright}
  \HRule
  \\[6mm]
  {\bf\Huge Machine learning for classification and quantification of monoclonal antibody    preparations for cancer therapy}
  \\[10mm]
  \parbox[b]{13cm}{\begin{flushright}
    {\Large
      Laetitia Le$^{~ab}$,
      Camille Marini$^{~cd}$,
      Alexandre Gramfort$^{~cef}$,
      David Nguyen$^{~a}$,
      Mehdi Cherti$^{~cg}$,
      Sana Tfaili$^{~b}$,
      Ali Tfayli$^{~b}$,
      Arlette Baillet-Guffroy $^{~b}$,
      Patrice Prognon $^{~ab}$,  
      Pierre Chaminade $^{~b}$,      
      Eric Caudron$^{~ab}$,
      Bal\'azs K\'egl$^{~cg}$
    }
    \\[6mm]
    \parbox[b]{10cm}{\begin{flushright}
     $^a$ European Georges Pompidou Hospital (AP-HP), Pharmacy department, Paris, France\\
     $^b$ Lip(Sys)² Chimi Analytique Pharmaceutique, Univ. Paris-Sud, Université Paris Saclay, F92290 Chatenay-Malabry, France (EA4041 Groupe de Chimie Analytique de Paris Sud)\\       	     
     $^c$ Center of Data Science, Universit\'e Paris-Saclay\\
     $^d$ CMAP, Ecole Polytechnique, Palaiseau, France\\
     $^e$ INRIA, Parietal team, Saclay, France\\
     $^f$ LTCI, T\'el\'ecom ParisTech\\
     $^g$ LAL, CNRS, France\\
    \end{flushright}}
    \\[5mm]
    26 March 2017\\ \end{flushright}
    }
  \\
  \HRule
\end{flushright}
\vspace*{\stretch{2}}

\centerline{\bf Abstract}

\vspace{12pt}

Monoclonal antibodies constitute one of the most important strategies to treat patients suffering from cancers such as hematological malignancies and solid tumors. In order to guarantee the quality of those preparations prepared at hospital, quality control has to be developed. The aim of this study was to explore a noninvasive, nondestructive, and rapid analytical method to ensure the quality of the final preparation without causing any delay in the process. We analyzed four mAbs (Inlfiximab, Bevacizumab, Ramucirumab and Rituximab) diluted at therapeutic concentration in chloride sodium 0.9\% using Raman spectroscopy. To reduce the prediction errors obtained with traditional chemometric data analysis, we explored a data-driven approach using statistical machine learning methods where preprocessing and predictive models are jointly optimized. We prepared a data analytics workflow and submitted the problem to a collaborative data challenge platform called Rapid Analytics and Model Prototyping (RAMP). This allowed to use solutions from about 500 data scientists during five days of collaborative work. The prediction of the four mAbs samples was considerably improved with a misclassification rate and the mean error rate of 0.8\% and 4\%, respectively.

\vspace*{\stretch{3}}

\clearpage

\tableofcontents

\clearpage
%
%

\section{Introduction}

Cancer can be treated using numerous strategies, such as surgery, radiation therapy, immunotherapy, hormone therapy, stem cell transplant, chemotherapy, and, recently, targeted therapy. Targeted therapy, for example, using monoclonal antibodies (mAbs), is the foundation of precision medicine. At present, it constitutes one of the most important strategies to treat patients suffering from cancers such as hematological malignancies and solid tumors. Monoclonal antibodies are proteins which bind to specific substance on cancer cells and act by immune-mediated cell killing mechanisms. They may therefore help the immune system to destroy cancer cells, stop cancer cells from growing, stop signals that help to form blood vessels, deliver cell-killing substances to cancer cells, and cause cancer cell death. As classic chemotherapy drugs, mAbs are commonly used alone or with other cytotoxic drugs or radioactive substances to kill cancer cells. 

Those drugs designed for parenteral administration are aseptically prepared by pharmacists. Drugs are extemporarily reconstituted or diluted in 5\% glucose or 0.9\% sodium chloride to obtain a sterile final preparation at the dose prescribed by the physician. Even if pharmaceutical regulation does not require final characterization of each compounding drugs, numerous pharmacists have nevertheless implemented analytical control. Those controls have to be discriminant to ensure the nature of the drug in spite of similar physicochemical and spectral properties, sensitive to guarantee the dose even for low concentrations and fast to secure the medication process without delayed drug delivery. Consequently, different analytical strategies using direct flow injection analysis, high performance liquid chromatography with UV detection~\citep{19362442}, or vibrational molecular spectroscopies such as Raman and infrared caudron~\citep{27769413, 24401426, 24463044} have been investigated to control cytotoxic compounding drugs. Despite their importance in the cancer therapy, only few studies reported on analytical methods to control mAbs preparations~\citep{20569773,27473490,27334717,25964136,22789912,24093128}. All of these techniques were invasive and required a sampling of the preparation for analysis. Because of the inherent toxicity of cytotoxic drugs as cancerogenic, mutagenic and teratogenic properties, these drugs present a risk of occupational exposure for healthcare workers. Thus, we decided to explore the feasibility of a noninvasive, nondestructive, and rapid analytical method as Raman spectroscopy to ensure the quality of the mAbs preparations produced at hospital. This molecular vibrational spectroscopy based on inelastic scattering of monochromatic light, usually from a laser in the visible, near infrared, or near ultraviolet range, was successfully investigated for chemotherapy drug control~\citep{27769413,24401426,25148728,24463044}. 

\begin{figure}[ht!]
\begin{center}
\includegraphics[width=0.70\columnwidth]{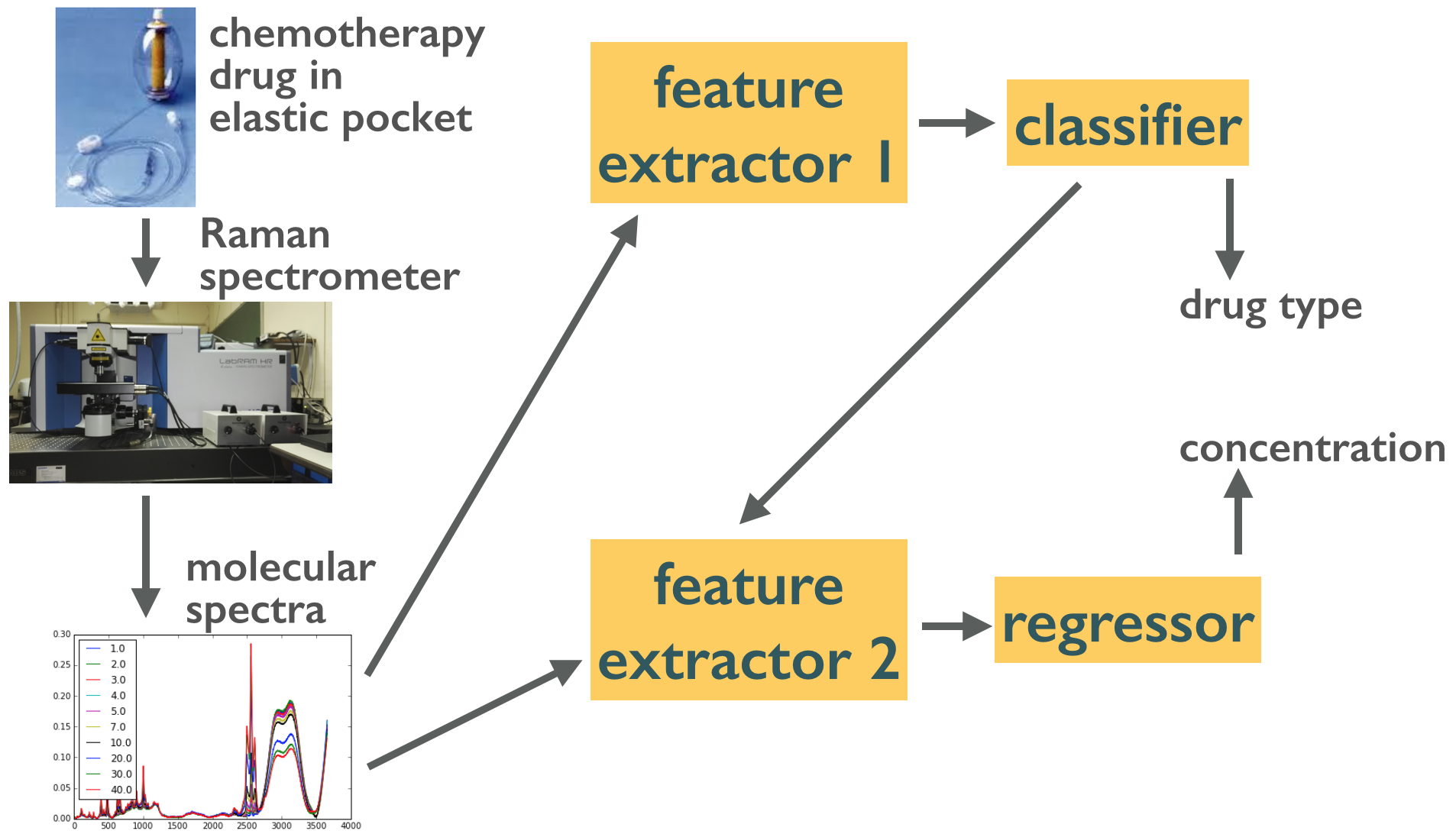}
\caption{\label{fig:workflow} The data acquisition and data analytics pipeline.%
}
\end{center}
\end{figure}

Due to the complexity of Raman spectral data, multivariate analysis has to be used to extract pertinent information. We have tried traditional linear techniques (partial least squares discriminant analysis and regression) and we obtained unsatisfactory results at both the molecule classification and concentration regression tasks. In this paper we report results using machine learning methods. We have built an analytics workflow (Figure~\ref{fig:workflow}), and submitted the data to a collaborative data challenge platform called Rapid Analytics and Model Prototyping (RAMP; \url{http://ramp.studio}), developed by the Paris-Saclay Center of Data Science. A large number of feature extraction and prediction techniques were submitted by about 500 data scientists. The top results, reported in this paper, were obtained after about five days of collaborative work.

\section{Material and methods}

\subsection{mAbs Samples}

We evaluated four mAbs: Bevacizumab (Avastin$^{\tiny{\text{\textregistered}}}$ \unit[25]{mg/mL} from Roche), Infliximab (Remicade$^{\tiny{\text{\textregistered}}}$ \unit[100]{mg} from Schering-Plough), Ramucirumab (Cyramza$^{\tiny{\text{\textregistered}}}$ \unit[10]{mg/mL} from Lilly), and Rituximab (Mabthera$^{\tiny{\text{\textregistered}}}$ \unit[10]{mg/mL} from Roche). We prepared all drugs separately in aseptic condition and analyzed them after dilution in 0.9\% chloride sodium at various concentrations that cover the therapeutic range currently prepared to treat patients (10 concentrations for each drug: Bevacizumab from 0.5 to \unit[25]{mg/mL}, Infliximab from 0.3 to \unit[10]{mg/mL}, Ramucirumab from 1 to \unit[10]{mg/mL} and Rituximab from 0.4 to \unit[10]{mg/mL}). We prepared 12 independent series (12 batch of 0.9\% chloride sodium) for each drug. We conditioned all compounded solutions into 3 glass vials (Interchim$^{\tiny{\text{\textregistered}}}$, Montluçon, France), conserved at \unit[+4]{$^\circ$C}, and analyzed in accordance to laboratory requirements.

Throughout the paper, we will use the following labels for the four drugs: A= Infliximab; B=Bevacizumab; Q=Ramucirumab, R=Rituximab.

\subsection{Raman spectroscopy}

We performed Raman spectral acquisitions with a Labram HR evolution microspectrometer (Horiba Jobin Yvon, Lille, France). The excitation source was a \unit[633]{nm} single-mode diode laser (Toptica Photonics, Germany) generating \unit[35]{mW} on the sample. The microspectrometer was equipped with an Olympus microscope and measurements were recorded using 10 and 100X MPlan objectives (Olympus, Japan). We collected light scattered by the sample through the same objective. We intercepted Rayleigh elastic scattering by an edge filter. A Peltier cooled (\unit[-70]{$^\circ$C}) multichannel CCD detector (Coupled Charge Device; 1024 $\times$ 256 pixels) detected the Raman Stokes signal dispersed with a \unit[100]{$\mu$m} slit width and \unit[600]{grooves/mm} holographic grating. The spectral resolution from the full width at half maximum of the silica wafer band at \unit[521]{cm$^{-1}$} was \unit[2]{cm$^{-1}$}. We studied the spectral region of \unit[400 - 4000]{cm$^{-1}$}. The acquisition time of each spectrum was \unit[2 $\times$ 15]{s} per collected spectrum to have better sensitivity. We conducted spectral acquisition and data pre-processing with Labspec6 software (Horiba Jobin Yvon SAS, Lille, France). Because the sample compartment was not adapted to analyze vials, we used a homemade vial adaptor to centralize the vial and to secure the position of the sample on the base plate. In order to correct spectral variation due to focalization, we pre-treated all Raman spectra using LabSpec6 by normalization based on total area. The result of this collection step was a spectrum containing $1866$ values.

\begin{figure}[h!]
\begin{center}
\centering
\subfigure[All spectra]{ 
\includegraphics[width=0.48\columnwidth]{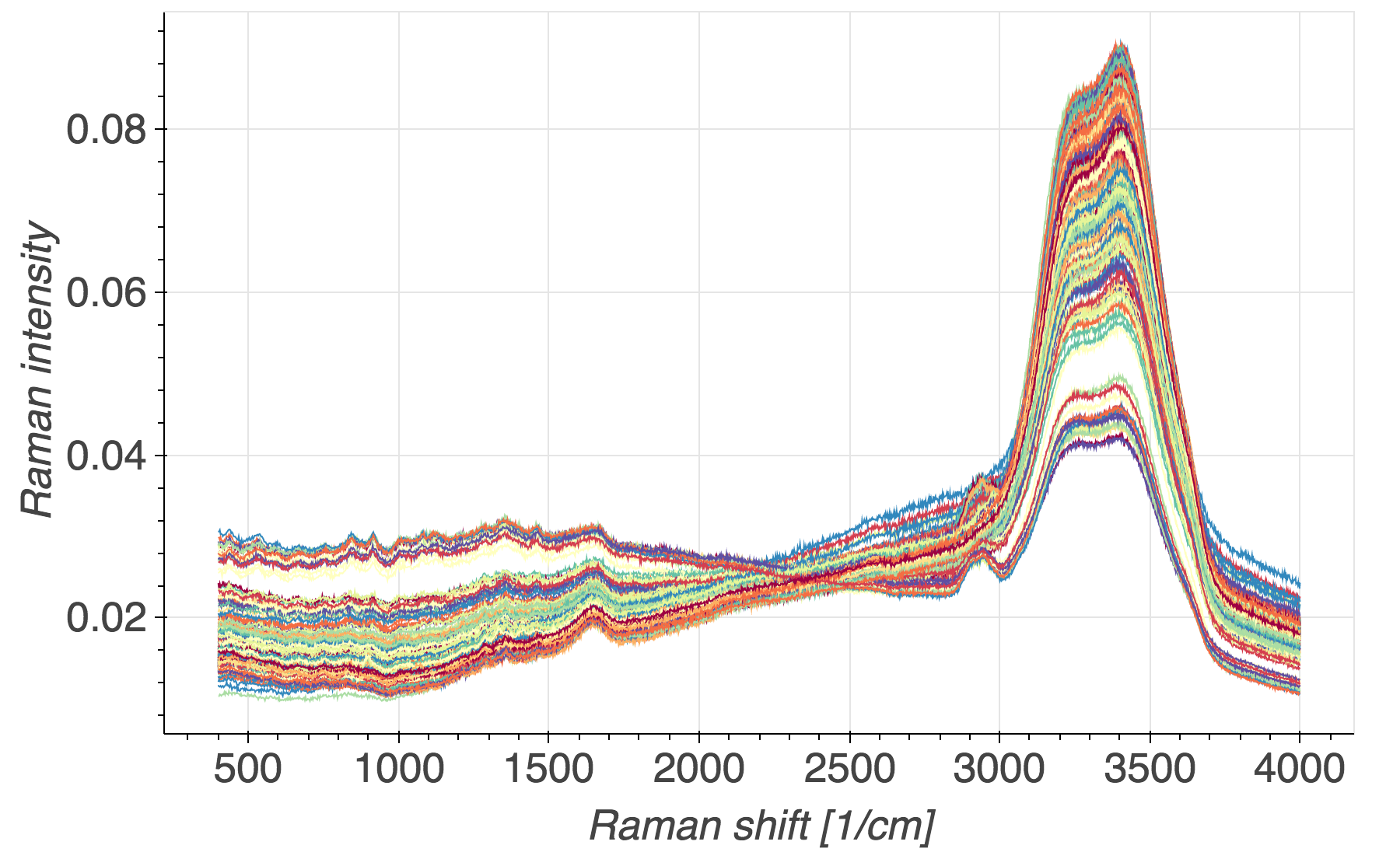}
\label{figAllSpectra}
}
\subfigure[Mean spectra per concentration]{
\includegraphics[width=0.48\columnwidth]{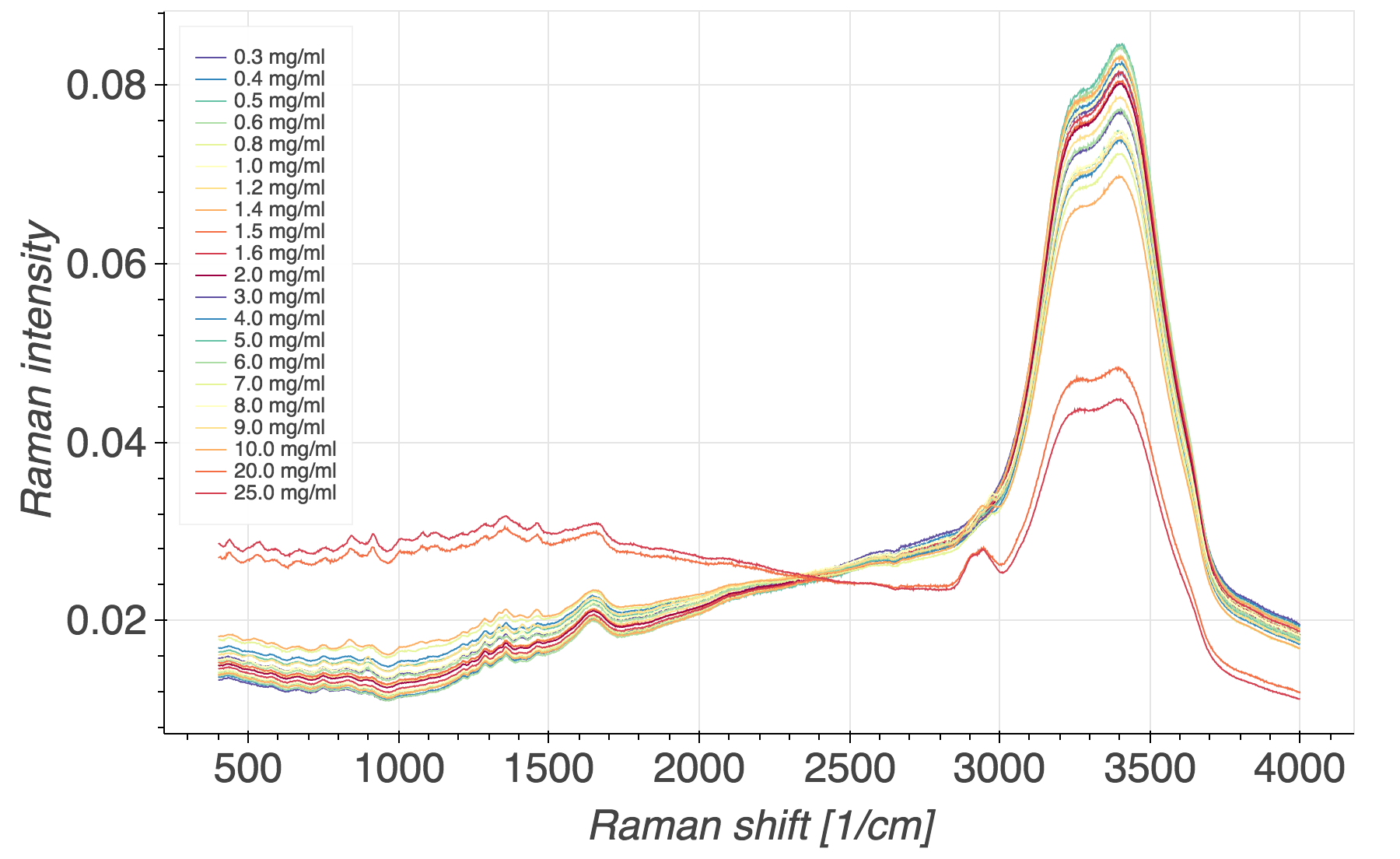}
\label{figSpectraPerConcentration}
}
\subfigure[Mean spectra per molecule]{
\includegraphics[width=0.48\columnwidth]{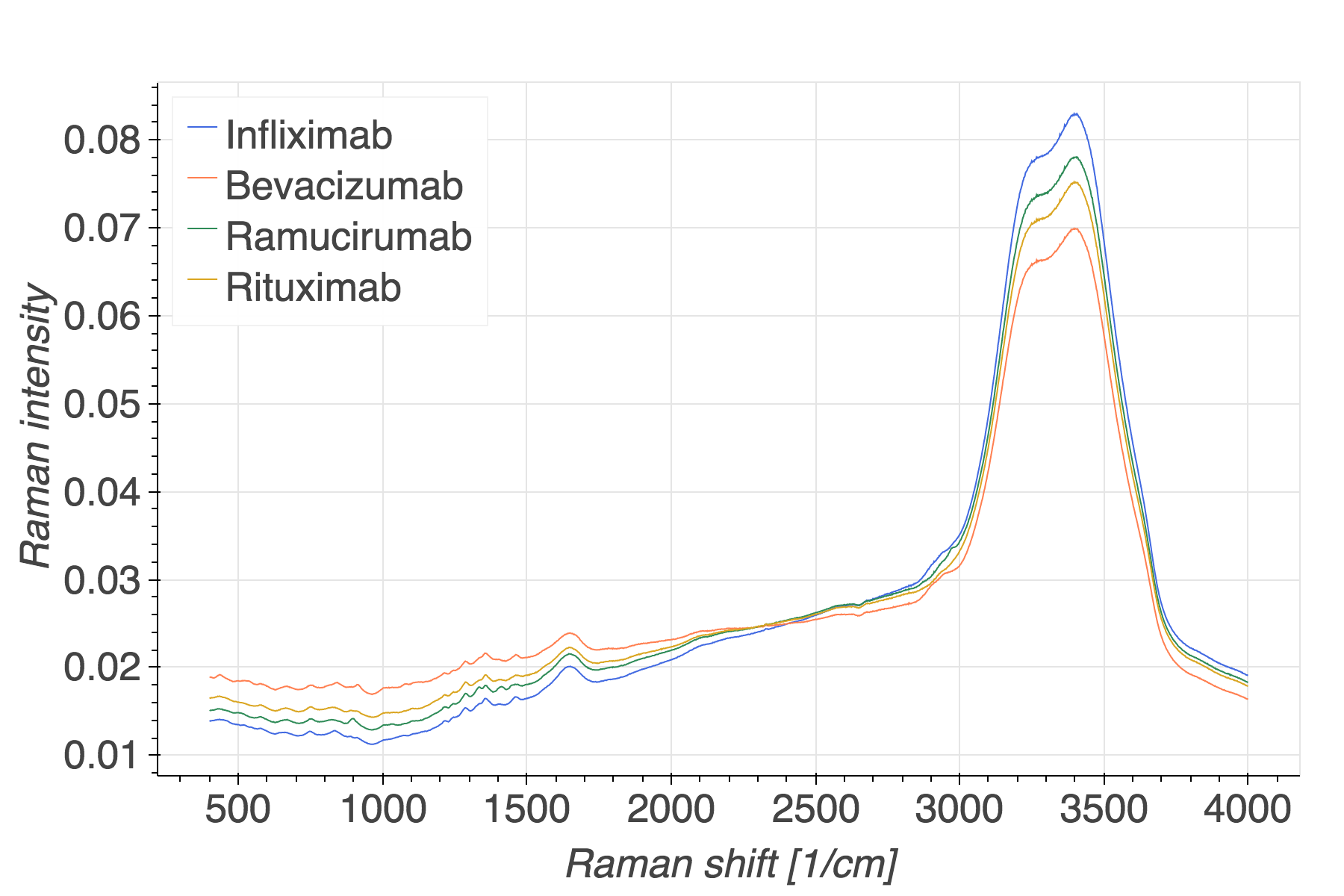}
\label{figSpectraPerConcentration}
}
\caption{\label{fig:spectra} Raw spectra and mean spectra grouped by different axes.%
}
\end{center}
\end{figure}

The raw spectra are shown in Figure~\ref{fig:spectra} for all concentrations and all molecules. Raman spectra of mAbs, similarly to that of other proteins, are very difficult to interpret. Inter and intra molecular effects including peptide-bond angles and hydrogen-bonding patterns may influence Raman band positions. Structural information can nevertheless be deduced from Raman vibrational bands as amide I and amide III bands at 1650 and \unit[1300]{cm$^{-1}$}, respectively.

\subsection{Data analysis}

In this section we formalize the data analysis problem and report overall results using traditional linear techniques and statistical machine learning approaches. We will present a more detailed analysis using the optimal technique in Section~\ref{secResults}.

The goal of the data analysis is to predict the drug type $\ell$ and its concentration $y$ based on the recorded Raman spectrum $\bs \in \RR^{1866}$. We used the following experimental protocol to develop and tune the analysis techniques and to test and compare the different methods. We collected a total of 360 spectra for each of the four mAbs agents, except for Ramucirumab (348 spectra). Following the standard machine learning protocol that consists in evaluating model performance on a held out set of data, the 1428 measurements were randomly split into two data sets. The 429 test instances were used to evaluate the developed techniques. All results reported in this paper are coming from this test set, but they were hidden from the analysts throughout development. The remaining 999 points were further split into eight random cross-validation folds. Each training and validation set contained 799 and 200 points, respectively. Each technique was trained eight times on the training sets. The eight trained models were then evaluated on each validation and test point. The final prediction is the average of these eight predictions on each validation and test point. This technique, commonly known as cross-validation bootstrap aggregation (CV bagging), takes advantage of the variance reduction property of averaging and allows us to develop robust predictors despite the small data size. To further improve the results, we \emph{blended} the best models using the technique of \citet{Caruana:2004:ESL:1015330.1015432}. In a nutshell, we take the pointwise mean prediction of the best models selected in a greedy loop until the validation result stops improving.

To allow the analysts to focus on various steps of the data analysis, we split the workflow into four modules (Figure~\ref{fig:workflow}). The first feature extractor $g^\text{clf}$ converts the raw spectra $\bs$ into a fixed size feature vector $\bx^\text{clf} = g^\text{clf}(\bs)$. This is then followed by a classifier $f^\text{clf}$ that outputs a vector $\bp = (p^\text{A}, p^\text{B}, p^\text{Q}, p^\text{R}) = f^\text{clf}\big(\bx^\text{clf}\big)$, indexed by the four drug types, representing the estimated probability that the spectrum belongs to each of the four types. The predicted class is then
\[
\widehat{\ell} = \argmax_{j \in \{\text{A, B, Q, R}\}} p^j \enspace .
\]
The second feature extractor $g^\text{reg}$ also receives the raw spectra $\bs$ but also the probability vector $\bc = f^\text{clf} (g^\text{clf} (\bs))$, and converts them into a fixed size feature vector $\bx^\text{reg} = g^\text{reg}(\bs, \bc)$. The rationale is to let the regression algorithm know the (estimated) identity of the molecule whose concentration it has to predict. The final module is a regression model $f^\text{reg}$ that takes this second feature vector $\bx^\text{reg}$ and estimates the drug concentration
\[
\widehat{y} = f^\text{reg}\big(\bx^\text{reg}\big) \enspace .
\]

Given a validation or test set $\cD = \big\{(\bs_i, \ell_i, y_i)\big\}_{i=1}^n$ containing triplets of spectra, molecule types, and concentrations, we evaluate the performance of the model $\cM = \big(g^\text{clf}, f^\text{clf}, g^\text{reg}, f^\text{reg}\big)$ as follows. We first compute the classification error
\[
R^\text{clf}(\cM, \cD) = \frac{1}{n} \sum_{i=1}^n \IND{\ell_i \not= \widehat{\ell}_i} \enspace ,
\]
where the indicator function $\IND{X}$ is one if its argument $X$ is true, and zero otherwise. To assess the concentration predictor, we use the mean absolute relative error (MARE)
\[
R^\text{reg}(\cM, \cD) = \frac{1}{n} \sum_{i=1}^n \frac{|y_i - \widehat{y}_i|}{y_i} \enspace .
\]
The analysts were instructed to minimize a combined error
\[
R^\text{comb}(\cM, \cD) = \frac{2}{3} R^\text{clf}(\cM, \cD) + \frac{1}{3} R^\text{reg}(\cM, \cD) \enspace .
\]
The coefficients are reflecting the more stringent requirements posed on the classification task.

\subsubsection{The traditional linear approach}
\label{secLinear}

To establish baseline performance, we applied linear discrimination and regression methods using the Matlab software R2011a, common in traditional chemometry. That is to say, both the preprocessing $g^\text{clf}$, $g^\text{reg}$ and the predictor functions $f^\text{clf}$, $f^\text{reg}$ were linear. Classification was performed using partial least squares discriminant analysis (PLS-DA) and principal component analysis discriminant analysis (PCA-DA). We used PLS regression for estimating the concentration. Both for classification and regression, we used various spectral processing techniques to limit the non-informative spectral background: baseline correction with a $n$\/th degree polynomial curve, first and second Savitsky-Golay derivatives, standard normal variate (SNV), and combined preprocessing. For each model, the optimal number of latent variables was determined by 10-fold cross validation.

The overall misclassification rate of the baseline method was $14.5\%$ while its MARE was $14.7\%$. More detailed analysis of the performance can be found in Section~\ref{secResults}.

\subsubsection{A crowdsourced machine learning approach}
\label{secML}

\begin{figure}[!ht]
\centering
\subfigure[The code submission interface]{ 
\includegraphics[width=0.45\columnwidth]{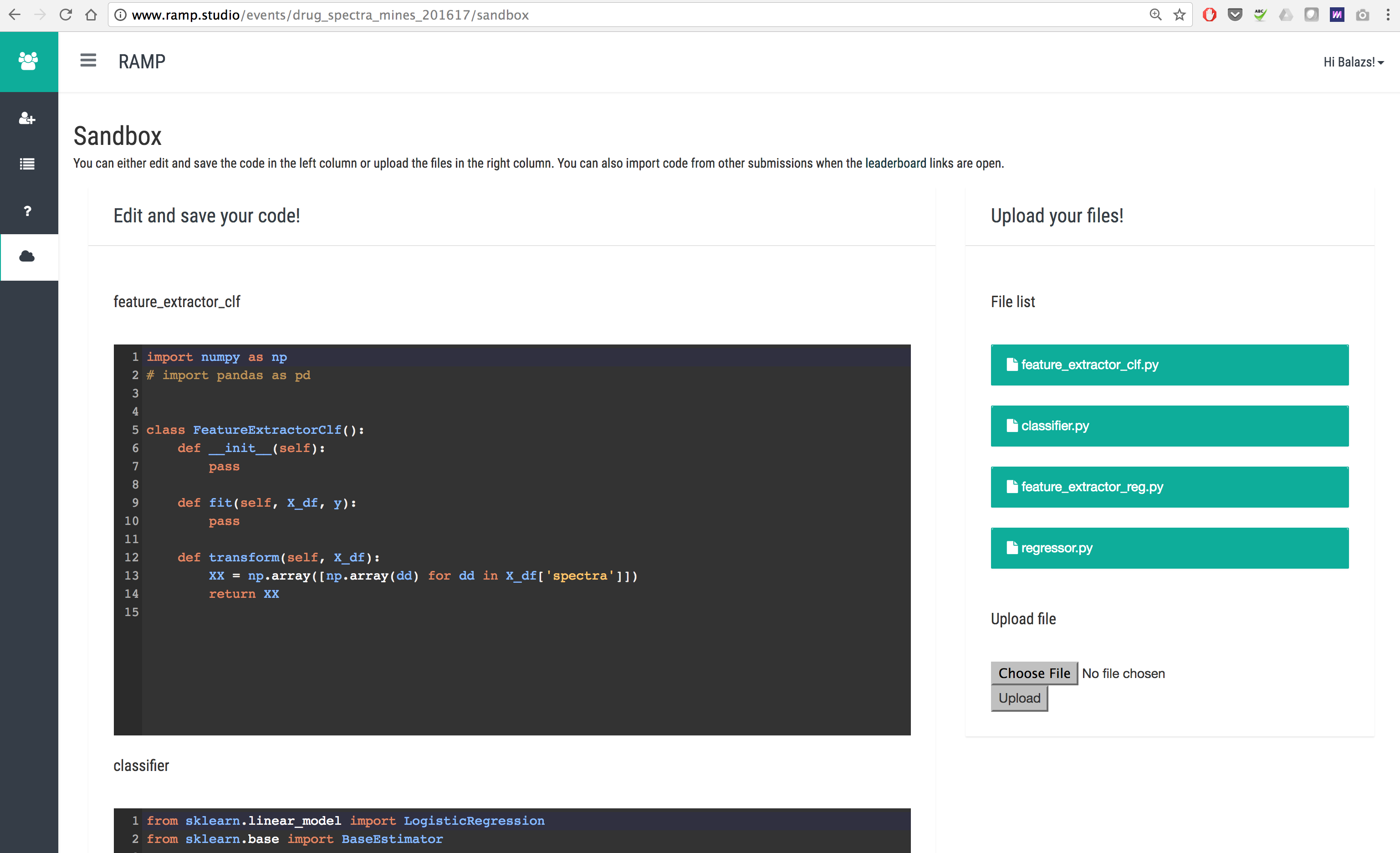}
\label{figSubmission}
}
\subfigure[The leaderboard]{
\includegraphics[width=0.45\columnwidth]{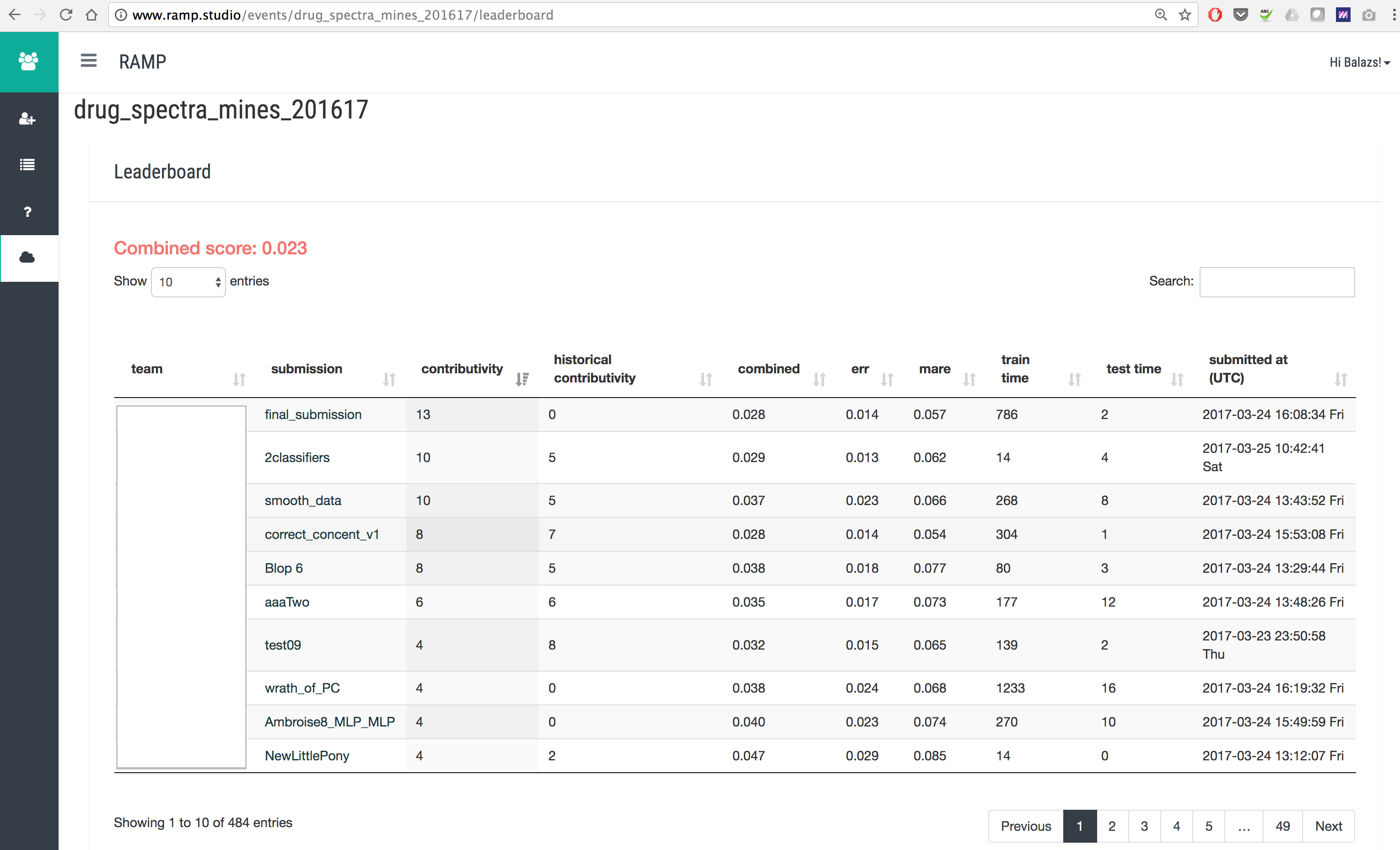}
\label{figLeaderboard}
}
\caption{(a) RAMP participants submit models (code) through a web interface and receive feedback through the (b) public leaderboard.}
\label{figRAMP}
\end{figure}

Machine learning is essentially a trial-and-error-based science. The \href{http://scikit-learn.org}{scikit-learn library} \citep{scikit-learn}, which most of our analysts used, contains dozens of different regression and classification techniques. Selecting and tuning the best predictor takes experience and a lot of experiments. The choice of feature extractors and filters is even larger. The \href{http://www.datascience-paris-saclay.fr}{Paris-Saclay Center for Data Science} has developed a unique \href{http://www.ramp.studio}{collaborative studio} which allows a large number of data scientists to collaborate on various scientific and industrial data analytics workflows. Participants have access to a brief description, the public training data set, and a first (non-optimized) solution in a starting kit. They study the problem, develop and tune solutions, and submit them through a web interface (Figure~\ref{figSubmission}). The models are then trained and evaluated, and the validation performance score is fed back to the participants through a public leaderboard (Figure~\ref{figLeaderboard}). The test score obtained on the held out test set, reported in this paper, remains hidden from the participants to avoid overfitting.

Unlike most of the data challenge sites that expect \emph{solutions} to analytics problems, we ask the data scientists to submit \emph{code}. This gives us more flexibility at the evaluation, generates a working prototype, and, most importantly, lets participants to collaborate by looking at each other's code and combining the different ideas. The code is publicly available at the \href{http://www.ramp.studio}{website} (after free sign-up), making the results fully reproducible by anyone. Moreover, any new method can be easily tested by submitting it on the site, so our results are fully transparent.

The drug spectra challenge was submitted to more than three hundred students in three different classroom RAMPs at Universit\'e Paris-Saclay, Ecole Polytechnique, and Mines ParisTech. Each RAMP involved about five days of work. In each class, we started with a closed period when students did not see each other's code, only their scores. This allowed us to grade the students individually but also to make them explore the space of solutions independently. This was followed by an open collaborative period. On the other hand, the classes could not see the solutions developed by the other classes. This, interestingly, led to different collective strategies in solving the problem. We blended the models in each RAMP separately which made it possible to compare the \emph{collaborative} score of each group.

The overall winner of the contest was the \href{http://www.ramp.studio/events/drug_spectra_mines_201617}{Mines RAMP} with a classification error of $0.7\%$ (only three low-concentration test samples were missed!) and a concentration MARE of $5.8\%$. The dominant solution that the group converged to was a log transformation of the spectra (but no other smoothing or preprocessing), followed by either \href{http://scikit-learn.org/stable/modules/generated/sklearn.decomposition.FactorAnalysis.html}{factor analysis} or \href{http://scikit-learn.org/stable/modules/generated/sklearn.decomposition.PCA.html}{principal component analysis} to extract 10 features, followed by a small \href{http://scikit-learn.org/stable/modules/generated/sklearn.neural_network.MLPClassifier.html}{neural network}. Interestingly, the same pipeline worked for both classification and regression, with, of course, different parameterizations of the neural net. In the regression step, all top models used the same strategy: to learn a different regression model (although parametrized the same way) for each molecule.

The \href{http://www.ramp.studio/events/drug_spectra}{Saclay RAMP} (chronologically the first) achieved a classification error of $1.6\%$ and a concentration MARE of $4.9\%$. This group did more preprocessing. Smoothing of the spectra was done with linear filters such as the \href{https://docs.scipy.org/doc/scipy-0.16.1/reference/generated/scipy.signal.savgol_filter.html}{Savitzky--Golay filter}~\citep{savgol} that allows to preserve the amplitudes of the peaks in the signal. Certain models used simpler strategies based on convolution with \href{https://docs.scipy.org/doc/numpy/reference/generated/numpy.hanning.html}{Hanning windows}. Following the smoothing step, the baseline correction was done by subtracting a polynomial least square fit of the data or simply by removing a constant. The polynomial order was 0 (constant) or 1 (linear). The order 0 corresponds to the subtraction of the mean of the spectra. Some of the solutions proposed to subtract the median of the spectra. As the machine learning models employed in the second step are sensitive to the scale of the data, the spectra, for some models, were normalized so that each spectrum after smoothing and baseline correction was of norm one. For prediction, the majority of winning solutions employed non-linear kernel-based techniques: \href{http://scikit-learn.org/stable/auto_examples/decomposition/plot_kernel_pca.html}{kernel PCA} for non-linear dimensionality reduction \citep{Scholkopf:1999} and \href{http://scikit-learn.org/stable/modules/svm.html}{support vector machines (SVM)} for prediction~\citep{Cristianini:1999:ISV:345662}. The best proposed approaches used Gaussian kernels and polynomial kernels of order up to 4. The low MARE was achieved by exploiting the fact that concentrations were discretized (see Section~\ref{secDiscussion}).

Finally, the \href{http://www.ramp.studio/events/drug_spectra}{Polytechnique RAMP} achieved a classification error of $2.1\%$ and a concentration MARE of $12.2\%$. The reason of the suboptimal performance was that the group was only exploring forest-based regression models (\href{http://scikit-learn.org/stable/modules/generated/sklearn.ensemble.ExtraTreesClassifier.html}{extra trees} \citep{Geurts:2006}, \href{http://scikit-learn.org/stable/modules/generated/sklearn.ensemble.RandomForestClassifier.html}{random forests} \citep{Breiman:2001}, \href{https://github.com/dmlc/xgboost}{gradient boosting} \citep{Friedman00greedyfunction}). These otherwise popular and usually well-performing nonparametric classifiers seemed to be a suboptimal choice for the functional data we had in this problem.

Note again that all the models and code are available freely at the \href{http://www.ramp.studio}{RAMP site}, so anybody can consult their detailed parameterizations, reuse them, and reproduce the results. Furthermore, anybody can resubmit competing solutions which will be tested using the same protocol as during the RAMPs, on the same training and test sets.

\section{Results}
\label{secResults}

Table~\ref{tab:RAMP_results} shows that the large improvement obtained by machine learning techniques is uniform across all molecules. The three misclassified molecules (that correspond to the overall misclassification rate of $0.7\%$) all came from samples with low concentration ($y \le \unit[0.6]{mg/ml}$). Using the linear method, $63\%$ of the samples below concentration $y \le \unit[1]{mg/ml}$ had larger relative error than $15\%$ whereas the figure was only $5.6\%$ in the case of the ML method.

\begin{table}
\begin{center}
  \begin{tabular}{|l|c|c|c|c| }
    \hline
    \textbf{molecule} & \textbf{miscl. rate (linear)} &
    \textbf{MARE (linear)}  & \textbf{miscl. rate (ML)} &
    \textbf{MARE (ML)} \\
    \hline
    Infliximab & $13.7\%$ & $12.3\%$ & $0.9\%$ & $8.4\%$\\
    \hline
    Bevacizumab & $19.8\%$ & $14.0\%$ & $1.0\%$ & $4.3\%$\\
    \hline
    Ramucirumab & $9.0\%$ & $7.3\%$ & $0.0\%$ & $3.5\%$\\
    \hline
    Rituximab & $16.0\%$ & $26.7\%$ & $1.0\%$ & $6.9\%$\\
    \hline
    \hline
    Overall & $14.5\%$ & $14.7\%$ & $0.7\%$ & $5.8\%$\\
    \hline
  \end{tabular}
\end{center}
\caption{\label{tab:RAMP_results} Misclassification rates and concentration MAREs obtained by the traditional chemometric (linear) baseline method (Section~\ref{secLinear}) and by the crowdsourced machine learning ensemble (Section~\ref{secML}) on 429 test points.}
\end{table}

\begin{figure}[h!]
\begin{center}
\centering
\subfigure[Bias of the concentration estimate]{ 
\includegraphics[width=0.48\columnwidth]{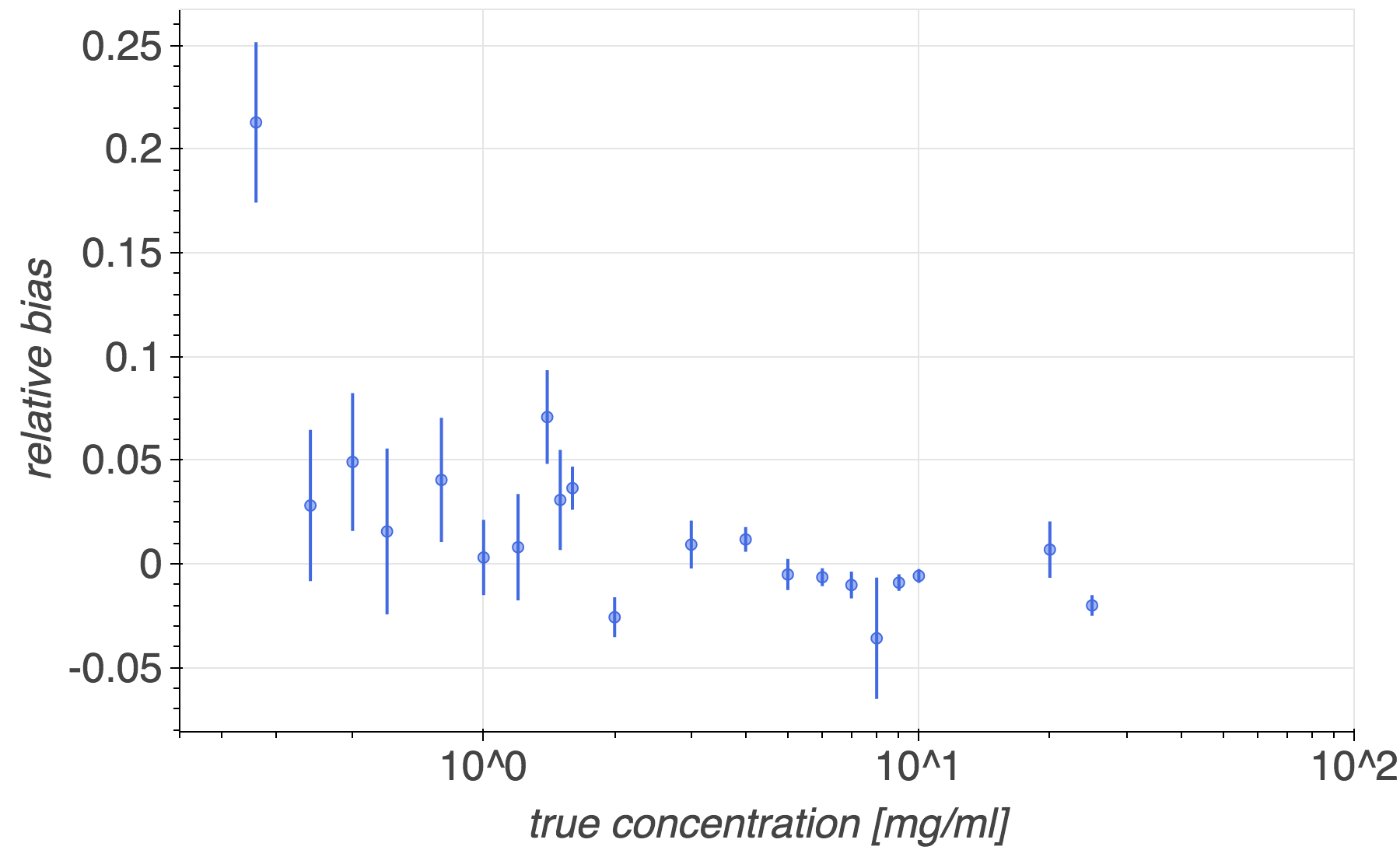}
\label{figConcentrationBias}
}
\subfigure[Resolution of the concentration estimate]{
\includegraphics[width=0.48\columnwidth]{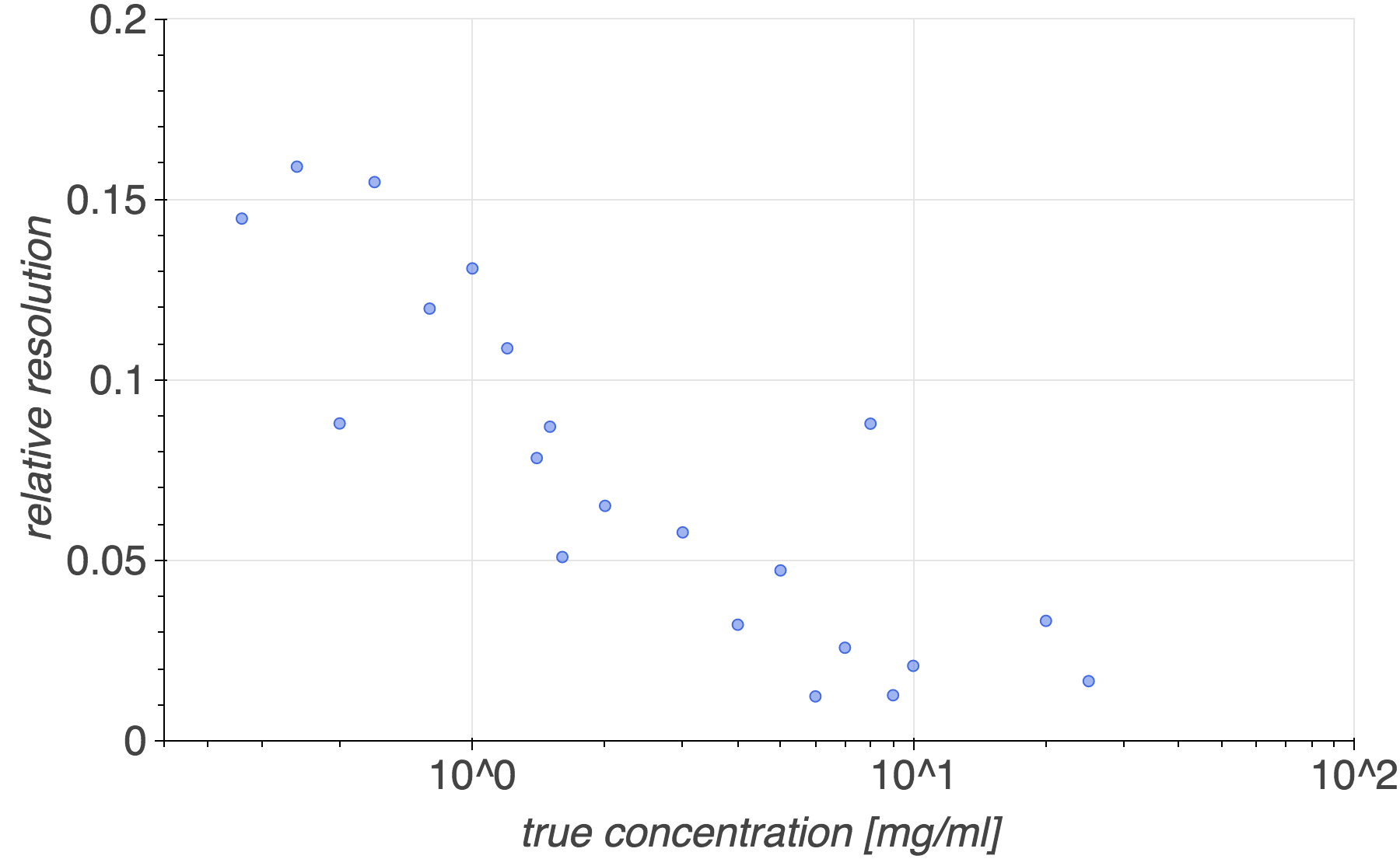}
\label{figConcentrationResolution}
}
\caption{{\label{figBiasResolution} The relative bias and resolution of the ML concentration estimate as a function of the concentration. Error bars are one standard deviation. %
}}
\end{center}
\end{figure}

Figure~\ref{figBiasResolution} shows the estimation bias (the mean of the relative error $(y - \widehat{y}) / y$) and the resolution (the standard deviation of the relative error) as a function of the concentration. Besides a downward slope explained by regression to the mean, the bias (Figure~\ref{figConcentrationBias}) is negligible above \unit[0.1]{mg/ml}. The relative resolution (Figure~\ref{figConcentrationResolution}) drops sharply with the concentration (not surprisingly) from 15\% (below $y < \unit[1]{mg/ml}$) to 5\% (above $y > \unit[3]{mg/ml}$).

\begin{figure}[h!]
\begin{center}
\includegraphics[width=0.7\columnwidth]{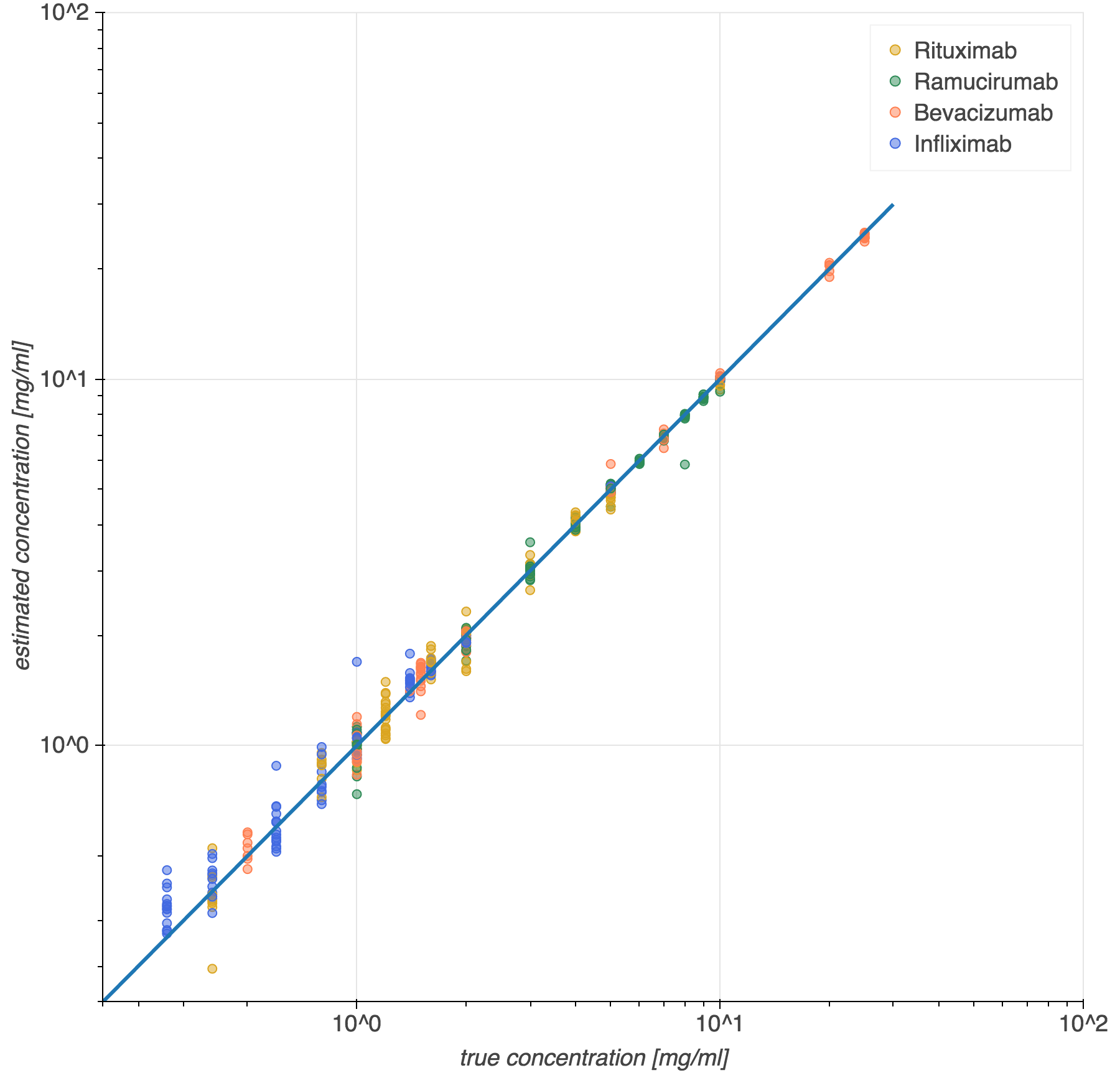}
\caption{{\label{fig:concentrations} True and predicted concentrations for each molecule. Predictions are obtained with the ensemble model of the \href{http://www.ramp.studio/events/drug_spectra_mines_201617}{Mines RAMP}.%
}}
\end{center}
\end{figure}

\section{Discussion}
\label{secDiscussion}

Over the last decade, the interest of statistical machine learning for biological applications has grown considerably. As part of both computer science and statistics, machine learning (ML), as a scientific discipline, aims to develop algorithms that can make sense of data. A typical outcome of machine learning is a model that can make prediction from new data, after having learnt from many training examples.
Based on the hypothesis that spectra are influenced by the nature and the concentration of mAbs in solution, we decided to explore how machine learning can help to resolve the pharmaceutical challenge of classification and concentration estimation for pharmaceutical drugs particularly difficult to discriminate by traditional analytical methods.

We found that combining Raman spectroscopy with machine learning methods present an interesting potential to secure the medication process by the identification and the quantification of chemotherapy preparations.

As opposed to traditional analysis using CLHP/UV or LC/MS/MS, Raman spectroscopies, such as near infrared spectroscopy, allow direct measurement through the glass and plastic packaging~\cite{11193093,12113892}. Because of rapid analysis and non-destructive and non-invasive measurements, those spectroscopies are largely used for the Process Analytical Technology~\cite{21167266} with at-line and on-line measurements to control primary and secondary manufacturing process in the pharmaceutical industry. 

Despite promising results to ensure the nature and the dose of the drug in solution by classification and regression analysis respectively, machine learning has some particular limitations and pitfalls that should be avoided. Machine learning algorithms usually require more training data to train than linear models. In addition, one has to be careful to collect data under the same distribution as when the predictors are used in practice. First, if the training concentrations are over- or under-sampled in certain regions, the nonparametric concentration estimate may be biased towards the oversampled values. This is due to the fact that machine learning models try to minimize the errors on average. Second, if only certain concentration levels are used for certain molecules, the classifier can learn these values and use the concentration information for classification. Third, if the concentration levels are discretized, the regression model can learn those discretization levels, and quantize the continuous estimate, creating a bias in case true concentrations fall between those levels. In fact, one of the student groups boosted the prediction accuracy by forcing the model to predict only concentrations present in the training set. More concretely, they rounded the predicted concentration by a regression model to the closest discrete concentration. In our study, the predictors were so precise that these effects were negligible, but in more complicated measurements, with larger uncertainties, they have to be taken into consideration. One practical way to consider in future data collection protocols in order to prevent performance gains by rounding is to randomize concentration values using a sampling distribution that matches realistic scenarios.

\section{Conclusion}

Due to similar structures and low concentration values, discrimination and quantification of mAbs preparations posed difficult challenges. In addition to the type of analytical technique, this study have highlighted the importance of the data analysis method and more particularly the power of statistical machine learning associated to Raman spectroscopy to ensure the chemical quality of medications produced at hospital before patient administration. 

\section*{Acknowledgement}

We would like to thank the 500 students and researchers participating in the initial hackaton and in the three classroom RAMPs. Their individual performance and credits can be assessed through their combined scores and contributivities at the three leaderboards (\href{http://www.ramp.studio/events/drug_spectra/leaderboard}{here}, \href{http://www.ramp.studio/events/drug_spectra_M1XMAP583_201617/leaderboard}{here}, and \href{http://www.ramp.studio/events/drug_spectra_mines_201617/leaderboard}{here}). We are also grateful for the Paris-Saclay Center for Data Science for supporting this research.

\bibliography{biblio}

\begin{thebibliography}{22}
\providecommand{\natexlab}[1]{#1}
\providecommand{\url}[1]{\texttt{#1}}
\expandafter\ifx\csname urlstyle\endcsname\relax
  \providecommand{\doi}[1]{doi: #1}\else
  \providecommand{\doi}{doi: \begingroup \urlstyle{rm}\Url}\fi

\bibitem[Amin et~al.(2014)Amin, Bourget, Vidal, and Ader]{25148728}
A.~Amin, P.~Bourget, F.~Vidal, and F.~Ader.
\newblock {Routine application of Raman spectroscopy in the quality control of
  hospital compounded ganciclovir.}
\newblock \emph{Int J Pharm}, 474:\penalty0 193--201, Oct 2014.

\bibitem[Ashton et~al.(2013)Ashton, Xu, Brewster, Cowcher, Sellick, Dickson,
  Stephens, and Goodacre]{24093128}
L.~Ashton, Y.~Xu, V.~Brewster, D.~Cowcher, C.~Sellick, A.~Dickson, G.~Stephens,
  and R.~Goodacre.
\newblock {The challenge of applying Raman spectroscopy to monitor recombinant
  antibody production.}
\newblock \emph{Analyst}, 138:\penalty0 6977--85, Nov 2013.

\bibitem[Bazin et~al.(2010)Bazin, Vieillard, Astier, and Paul]{20569773}
C.~Bazin, V.~Vieillard, A.~Astier, and M.~Paul.
\newblock {[Reliable real-time analytical control of monoclonal antibodies
  chemotherapies preparations on Multispec automaton].}
\newblock \emph{Ann Pharm Fr}, 68:\penalty0 163--77, May 2010.

\bibitem[Bourget et~al.(2014)Bourget, Amin, Vidal, Merlette, and
  Lagarce]{24463044}
P.~Bourget, A.~Amin, F.~Vidal, C.~Merlette, and F.~Lagarce.
\newblock {Comparison of Raman spectroscopy vs. high performance liquid
  chromatography for quality control of complex therapeutic objects: model of
  elastomeric portable pumps filled with a fluorouracil solution.}
\newblock \emph{J Pharm Biomed Anal}, 91:\penalty0 176--84, Mar 2014.

\bibitem[Breiman(2001)]{Breiman:2001}
L.~Breiman.
\newblock Random forests.
\newblock \emph{Mach. Learn.}, 45\penalty0 (1):\penalty0 5--32, Oct. 2001.
\newblock ISSN 0885-6125.
\newblock \doi{10.1023/A:1010933404324}.

\bibitem[Broad et~al.(2000)Broad, Jee, Moffat, Eaves, Mann, and
  Dziki]{11193093}
N.~Broad, R.~Jee, A.~Moffat, M.~Eaves, W.~Mann, and W.~Dziki.
\newblock {Non-invasive determination of ethanol, propylene glycol and water in
  a multi-component pharmaceutical oral liquid by direct measurement through
  amber plastic bottles using Fourier transform near-infrared spectroscopy.}
\newblock \emph{Analyst}, 125:\penalty0 2054--8, Nov 2000.

\bibitem[Caruana et~al.(2004)Caruana, Niculescu-Mizil, Crew, and
  Ksikes]{Caruana:2004:ESL:1015330.1015432}
R.~Caruana, A.~Niculescu-Mizil, G.~Crew, and A.~Ksikes.
\newblock {Ensemble Selection from Libraries of Models}.
\newblock In \emph{Proceedings of the Twenty-first International Conference on
  Machine Learning}, ICML '04, pages 18--, New York, NY, USA, 2004. ACM.
\newblock ISBN 1-58113-838-5.
\newblock \doi{10.1145/1015330.1015432}.

\bibitem[Cristianini and Shawe-Taylor(2000)]{Cristianini:1999:ISV:345662}
N.~Cristianini and J.~Shawe-Taylor.
\newblock \emph{{An Introduction to Support Vector Machines: And Other
  Kernel-based Learning Methods}}.
\newblock Cambridge University Press, New York, NY, USA, 2000.
\newblock ISBN 0-521-78019-5.

\bibitem[De et~al.(2011)De, Burggraeve, Fonteyne, Saerens, Remon, and
  Vervaet]{21167266}
B.~T. De, A.~Burggraeve, M.~Fonteyne, L.~Saerens, J.~Remon, and C.~Vervaet.
\newblock {Near infrared and Raman spectroscopy for the in-process monitoring
  of pharmaceutical production processes.}
\newblock \emph{Int J Pharm}, 417:\penalty0 32--47, Sep 2011.

\bibitem[Delmas et~al.(2009)Delmas, Gordien, Bernadou, Roudaut, Gresser, Malki,
  Saux, and Breilh]{19362442}
A.~Delmas, J.~Gordien, J.~Bernadou, M.~Roudaut, A.~Gresser, L.~Malki, M.~Saux,
  and D.~Breilh.
\newblock {Quantitative and qualitative control of cytotoxic preparations by
  HPLC-UV in a centralized parenteral preparations unit.}
\newblock \emph{J Pharm Biomed Anal}, 49:\penalty0 1213--20, Jul 2009.

\bibitem[Friedman(2000)]{Friedman00greedyfunction}
J.~H. Friedman.
\newblock Greedy function approximation: A gradient boosting machine.
\newblock \emph{Annals of Statistics}, 29:\penalty0 1189--1232, 2000.

\bibitem[Geurts et~al.(2006)Geurts, Ernst, and Wehenkel]{Geurts:2006}
P.~Geurts, D.~Ernst, and L.~Wehenkel.
\newblock Extremely randomized trees.
\newblock \emph{Mach. Learn.}, 63\penalty0 (1):\penalty0 3--42, Apr. 2006.
\newblock ISSN 0885-6125.
\newblock \doi{10.1007/s10994-006-6226-1}.

\bibitem[Jaccoulet et~al.(2015)Jaccoulet, Smadja, Prognon, and
  Taverna]{25964136}
E.~Jaccoulet, C.~Smadja, P.~Prognon, and M.~Taverna.
\newblock {Capillary electrophoresis for rapid identification of monoclonal
  antibodies for routine application in hospital.}
\newblock \emph{Electrophoresis}, 36:\penalty0 2050--6, Sep 2015.

\bibitem[Jaccoulet et~al.(2016{\natexlab{a}})Jaccoulet, Boccard, Taverna,
  Azevedos, Rudaz, and Smadja]{27334717}
E.~Jaccoulet, J.~Boccard, M.~Taverna, A.~Azevedos, S.~Rudaz, and C.~Smadja.
\newblock {High-throughput identification of monoclonal antibodies after
  compounding by UV spectroscopy coupled to chemometrics analysis.}
\newblock \emph{Anal Bioanal Chem}, 408:\penalty0 5915--24, Aug
  2016{\natexlab{a}}.

\bibitem[Jaccoulet et~al.(2016{\natexlab{b}})Jaccoulet, Smadja, and
  Taverna]{27473490}
E.~Jaccoulet, C.~Smadja, and M.~Taverna.
\newblock {Quality Control of Therapeutic Monoclonal Antibodies at the Hospital
  After Their Compounding and Before Their Administration to Patients.}
\newblock \emph{Methods Mol Biol}, 1466:\penalty0 179--84, 2016{\natexlab{b}}.

\bibitem[L\^{e} et~al.(2014)L\^{e}, Caudron, Baillet-Guffroy, and
  Eveleigh]{24401426}
L.~L\^{e}, E.~Caudron, A.~Baillet-Guffroy, and L.~Eveleigh.
\newblock {Non-invasive quantification of 5 fluorouracil and gemcitabine in
  aqueous matrix by direct measurement through glass vials using near-infrared
  spectroscopy.}
\newblock \emph{Talanta}, 119:\penalty0 361--6, Feb 2014.

\bibitem[L\^{e} et~al.(2016)L\^{e}, Tfayli, Zhou, Prognon, Baillet-Guffroy, and
  Caudron]{27769413}
L.~L\^{e}, A.~Tfayli, J.~Zhou, P.~Prognon, A.~Baillet-Guffroy, and E.~Caudron.
\newblock {Discrimination and quantification of two isomeric antineoplastic
  drugs by rapid and non-invasive analytical control using a handheld Raman
  spectrometer.}
\newblock \emph{Talanta}, 161:\penalty0 320--324, Dec 2016.

\bibitem[Paul et~al.(2012)Paul, Vieillard, Jaccoulet, and Astier]{22789912}
M.~Paul, V.~Vieillard, E.~Jaccoulet, and A.~Astier.
\newblock {Long-term stability of diluted solutions of the monoclonal antibody
  rituximab.}
\newblock \emph{Int J Pharm}, 436:\penalty0 282--90, Oct 2012.

\bibitem[Pedregosa et~al.(2011)Pedregosa, Varoquaux, Gramfort, Michel, Thirion,
  Grisel, Blondel, Prettenhofer, Weiss, Dubourg, Vanderplas, Passos,
  Cournapeau, Brucher, Perrot, and Duchesnay]{scikit-learn}
F.~Pedregosa, G.~Varoquaux, A.~Gramfort, V.~Michel, B.~Thirion, O.~Grisel,
  M.~Blondel, P.~Prettenhofer, R.~Weiss, V.~Dubourg, J.~Vanderplas, A.~Passos,
  D.~Cournapeau, M.~Brucher, M.~Perrot, and E.~Duchesnay.
\newblock {Scikit-learn: Machine Learning in {P}ython}.
\newblock \emph{Journal of Machine Learning Research}, 12:\penalty0 2825--2830,
  2011.

\bibitem[Schafer(2011)]{savgol}
R.~W. Schafer.
\newblock {What Is a Savitzky-Golay Filter?}
\newblock \emph{IEEE Signal Processing Magazine}, 28\penalty0 (4):\penalty0
  111--117, July 2011.
\newblock ISSN 1053-5888.
\newblock \doi{10.1109/MSP.2011.941097}.

\bibitem[Sch\"{o}lkopf et~al.(1998)Sch\"{o}lkopf, Smola, and
  M\"{u}ller]{Scholkopf:1999}
B.~Sch\"{o}lkopf, A.~Smola, and K.-R. M\"{u}ller.
\newblock Nonlinear component analysis as a kernel eigenvalue problem.
\newblock \emph{Neural Comput.}, 10\penalty0 (5):\penalty0 1299--1319, July
  1998.
\newblock ISSN 0899-7667.
\newblock \doi{10.1162/089976698300017467}.

\bibitem[Vergote et~al.(2002)Vergote, Vervaet, Remon, Haemers, and
  Verpoort]{12113892}
G.~Vergote, C.~Vervaet, J.~Remon, T.~Haemers, and F.~Verpoort.
\newblock {Near-infrared FT-Raman spectroscopy as a rapid analytical tool for
  the determination of diltiazem hydrochloride in tablets.}
\newblock \emph{Eur J Pharm Sci}, 16:\penalty0 63--7, Jul 2002.

\end{thebibliography}

\end{document}